\documentclass[secnumarabic,amssymb,nobibnotes,aps,prl,preprint,letterpaper,preprintnumbers,amsmath,superscriptaddress,showpacs,floatfix]{revtex4-1}

\usepackage{wrapfig}
\usepackage{amsmath}
\usepackage{enumitem}
\usepackage{multirow}
\usepackage[letterpaper,margin={1in,1in}]{geometry}
\usepackage{graphicx}
\usepackage{lineno}
\usepackage{appendix}
\usepackage{longtable}
\usepackage{array}
\usepackage[dvipsnames]{xcolor}

\usepackage{multirow}

\begin{document}
\setlength{\LTcapwidth}{6.5in}

\title{Infrared nano-spectroscopy of ferroelastic domain walls in  hybrid improper ferroelectric Ca$_3$Ti$_2$O$_7$}

\author{K. A. Smith}
\affiliation{Department of Chemistry, University of Tennessee,
Knoxville, Tennessee 37996 USA}

\author{E. A. Nowadnick}
\affiliation{Department of Physics, New Jersey Institute of Technology, Newark, NJ 07102 USA}
\affiliation{School of Applied and Engineering Physics, Cornell
University, Ithaca, NY 14850 USA}
\affiliation{Department of Materials Science and Engineering
University of California, Merced, Merced, CA 95343}

\author{S. Fan}
\affiliation{Department of Physics, University of Tennessee,
Knoxville, Tennessee 37996 USA}

\author{O. Khatib}
\affiliation{Department of Physics, Department of Chemistry, and
JILA, University of Colorado, Boulder, Colorado 80309 USA}
\affiliation{Advanced Light Source Division, Lawrence Berkeley
National Laboratory, Berkeley, CA 94720 USA}

\author{S. J. Lim}
\affiliation{Rutgers Center for Emergent Materials, Rutgers
University, Piscataway, New Jersey 08854 USA}
\affiliation{Department of Physics and Astronomy, Rutgers University,
Piscataway, New Jersey 08854 USA}

\author{B. Gao}
\affiliation{Department of Physics and Astronomy, Rutgers University,
Piscataway, New Jersey 08854 USA}

\author{N. C. Harms}
\affiliation{Department of Chemistry, University of Tennessee,
Knoxville, Tennessee 37996 USA}

\author{S. N. Neal}
\affiliation{Department of Chemistry, University of Tennessee,
Knoxville, Tennessee 37996 USA}

\author{J. K. Kirkland}
\affiliation{Department of Chemistry, University of Tennessee,
Knoxville, Tennessee 37996 USA}

\author{M. C. Martin}
\affiliation{Advanced Light Source Division, Lawrence Berkeley
National Laboratory, Berkeley, CA 94720 USA}

\author{C. J. Won}
\affiliation{Laboratory for Pohang Emergent Materials, Pohang Accelerator Laboratory and Max Planck POSTECH Center for Complex Phase Materials, Pohang University of Science and Technology, Pohang 790-784, Korea}

\author{M. B. Raschke}
\affiliation{Department of Physics, Department of Chemistry, and
JILA, University of Colorado, Boulder, Colorado 80309 USA}

\author{S. -W. Cheong}
\affiliation{Rutgers Center for Emergent Materials, Rutgers
University, Piscataway, New Jersey 08854 USA}
\affiliation{Department of Physics and Astronomy, Rutgers University,
Piscataway, New Jersey 08854 USA}
\affiliation{Laboratory for Pohang Emergent Materials, Pohang Accelerator Laboratory and Max Planck POSTECH Center for Complex Phase Materials, Pohang University of Science and Technology, Pohang 790-784, Korea}

\author{C. J. Fennie}
\affiliation{School of Applied and Engineering Physics, Cornell
University, Ithaca, NY 14850 USA}

\author{G. L. Carr}
\affiliation{National Synchrotron Light Source II, Brookhaven National
Laboratory, Upton, New York 11973 USA}

\author{H. A. Bechtel}
\affiliation{Advanced Light Source Division, Lawrence Berkeley
National Laboratory, Berkeley, CA 94720 USA}

\author{J. L. Musfeldt}
\affiliation{Department of Chemistry, University of Tennessee,
Knoxville, Tennessee 37996 USA} \affiliation{Department of Physics,
University of Tennessee, Knoxville, Tennessee 37996 USA}

\date{\today}

\begin{abstract}

{\bf

Ferroic materials are well known to exhibit heterogeneity in the form of domain walls.  
Understanding the properties of these boundaries is crucial for controlling functionality with external stimuli and for realizing their potential  for ultra-low power memory and logic devices as well as novel computing architectures. 
In this work, we employ synchrotron-based near-field infrared nano-spectroscopy to reveal the vibrational properties of ferroelastic (90$^{\circ}$ ferroelectric) domain walls in the hybrid improper ferroelectric Ca$_3$Ti$_2$O$_7$. By locally mapping the  Ti-O stretching and Ti-O-Ti  bending modes, we reveal how structural order parameters rotate across a wall.  Thus, we link observed near-field amplitude changes  to underlying structural modulations and test ferroelectric switching models against real space measurements of local structure. This initiative opens the door to broadband infrared nano-imaging of heterogeneity in ferroics.}

\end{abstract}

\maketitle

\section{Introduction}

Many unique properties in functional materials arise from spatially heterogeneous electronic and magnetic states. Examples include the formation of charge and spin stripes in some cuprates and nickelates and phase separation across the Mott transition of VO$_2$    \cite{Tranquada1995,Dagotto2005,Katakura2010,Bhalla2009,Liu2013,Liu2015b,deQuilettes2015}. 
Ferroics also have naturally occurring heterogeneities in the form of domains and domain walls  
\cite{Scherpereel1970,Seidel2009,Choi2010,
VanAert2012,Meier2012,Schroder2012,
Yang22014,
Ma2015,Stolichnov2015,
Fiebig2016,Mundy2017}. 
This is
especially true for Ca$_3$Ti$_2$O$_7$, a hybrid improper  
ferroelectric 
\cite{Oh2015,Huang2016} where the 
polarization arises  from a trilinear coupling mechanism
\cite{Benedek2011, Harris2011,Benedek2015} and abundant
charged  domain walls have been observed \cite{Oh2015,Huang2016,Nowadnick2016}.
%
Atomic- and piezo-force imaging reveal the different orientations of
directional order parameters and  domain wall character,
providing a physical playground for graph
theory.
%
Like other counterpart materials \cite{Meier2012,Mundy2017}, ferroelectric domain walls in Ca$_3$Ti$_2$O$_7$ are  anisotropic and more conducting than their surroundings - although the domains themselves are insulating  \cite{Oh2015,Huang2016,Nowadnick2016}. 
Theoretical modeling and electron diffraction experiments recently revealed that polarization rotates across the ferroelectric walls 
in a manner that is more N\'eel- than Ising-like \cite{Lee2017}.  
%
Ferroelastic 
walls in Ca$_3$Ti$_2$O$_7$  are linear and 
separate different structural twin domains  \cite{Oh2015}. Because the boundaries connect domains with different spontaneous strain states, 
mechanical compatibility conditions determine their orientation \cite{Tagantsev2010,Nowadnick2016,Munro2018}.  These ferroelastic walls are also 90$^{\circ}$ ferroelectric walls. Within each twin, there also are meandering 180$^{\circ}$ ferroelectric walls \cite{Oh2015}.

 There is considerable interest in the dynamics of domain walls. 
 Broadband  scanning impedance microscopy of hexagonal ErMnO$_3$ 
 reveals a domain wall response  dominated by bound charge oscillations rather than free carrier conduction  \cite{Wu2017}.  Second harmonic generation spectroscopies  also provide insight into local symmetries and chemical structure, and at the same time, are sensitive 
imaging tools \cite{Fiebig2016}. 
X-ray photon correlation spectroscopy has uncovered ferroelectric domain wall dynamics in PbZr$_{0.55}$Ti$_{0.45}$O$_3$ as well \cite{Gorfman2018}.  
Despite a number of significant discoveries, almost nothing is known about  domain wall phonons. Infrared spectroscopy, a powerful tool for probing phonons and their symmetries, has been unable to address the nanoscale heterogeneities of domain walls due to its poor spatial resolution. The long wavelengths of infrared light necessarily mean that even for a diffraction-limited beam, traditional infrared techniques average over any micro- and nano-structured character of a target material, thereby reducing  the sensitivity to these heterogeneities \cite{Sun2014}. 
The development of near field spectroscopic techniques has made real space imaging of nanoscale heterogeneity  feasible, although most efforts are confined to the middle infrared due to the availability of suitable laser and broadband sources \cite{Huth2012,Muller2015}. 
%
%
%
The  extension into the far infrared using an accelerator-based source 
\cite{Omar} enables broad band measurement of the low-frequency response including phonons. 
Synchrotron-based near-field infrared spectroscopy \cite{Bechtel2014} thus has the potential to unlock the properties of defect states such as domain walls in complex oxides and chalcogenides. 
Near-field infrared spectroscopy is, of course, quite different than tip-enhanced Raman scattering in that ungerade (odd symmetry) phonon modes, which are crucial to the development of ferroelectricity and other functionalities, can be probed 
and analyzed.


 In order to uncover the behavior of fundamental excitations like
 phonons at domain walls and to explore the structural distortions
 that they represent, 
we performed far infrared synchrotron-based near-field nano-spectroscopy 
 of the local phonon response at domain boundaries in Ca$_3$Ti$_2$O$_7$ and compared  our  findings with theoretical
 models of how the order parameter evolves across the wall.
 %
 %
 %
 Analysis of the Ti-O stretching and bending modes across the
 ferroelastic 
 wall reveals subtle changes in the frequency and matrix element
 that we relate to the underlying modulation of the crystal structure and to the rotation of the
 structural order parameters.   There
 is significant width to the residual structural distortion across
 the walls (60 - 100 nm) and semiconducting character. 
 %
 This research opens the door to broadband imaging of heterogeneity in ferroics and  represents a first step to revealing the rich dynamics of domain walls in these systems. At the same time, it provides crucial information 
for the development of ultra low-power devices,  switches, polarizers, and computing
architectures based upon domain walls \cite{Mittall2016,Royo2017}. 
Loss mechanisms involving phonons are also key to controlling decoherence in  domain wall-based computing architectures \cite{Wu2017,Seijas-Bellido2017}. 
%

\section{Results and Discussion}
\subsection{Structure and order parameters across the domain boundary}

\begin{figure}[t!]
    \includegraphics[width=0.95\textwidth]{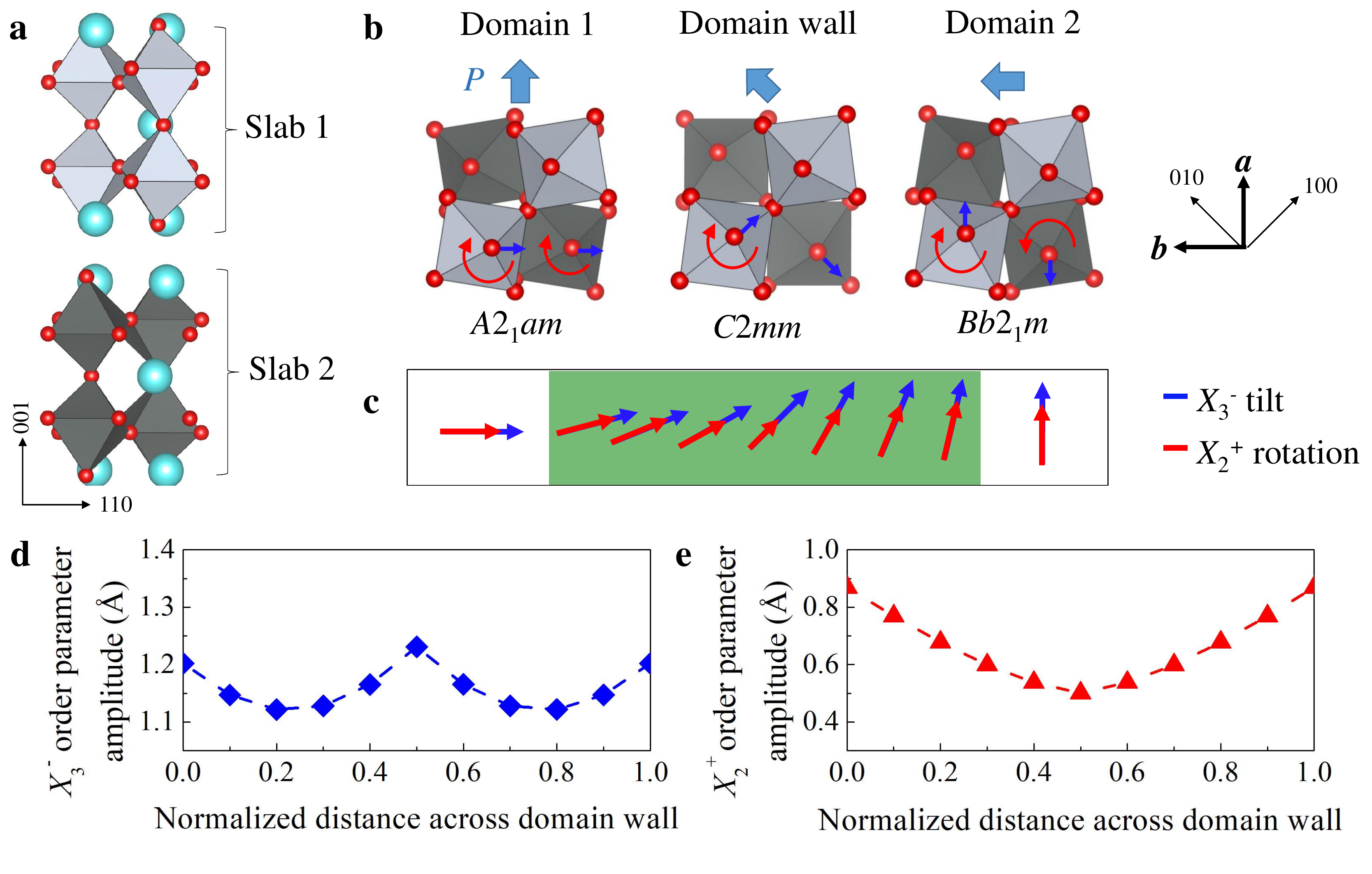}
    \caption{{\bf Bulk crystal and ferroelastic domain wall structure.} (a) Crystal structure of Ca$_3$Ti$_2$O$_7$ (space group $A2_1am$). The two CaTiO$_3$ perovskite slabs are shown in light and dark grey, with the Ca cations in light blue. The structure of the two orthorhombic twin domains is shown in (b) real space and (c) order parameter space. The order parameters for the $X_3^-$ octahedral tilt and $X_2^+$ octahedral rotation are shown with blue and red arrows, respectively, in (c) and the corresponding atomic motions are highlighted using the same colors in (b). The bold black arrows show the setting of the orthorhombic relative to the tetragonal axes. 
    The rotation direction  of the $\Gamma_5^-$ order parameter is indicated by the arrows with a $P$, indicating polarization direction, above the structures.  Upon crossing a ferroelastic domain wall, the octahedral rotation order parameters rotate by 90$^\circ$; at the midpoint, the local structure is space group $C2mm$. (d, e) The order parameter amplitudes  are plotted as a function of the normalized wall width. These amplitudes (reported for a $Z$ = 2 cell) were obtained using Density Functional Theory calculations of a trajectory through the bulk energy surface. The units of these structural order parameters are distance (in \AA) because they are measuring the amount of distortion - which is calculated by adding up how much all of the atoms in the unit cell are displaced from their high symmetry positions. See Supplemental Information for details. 
    }
    \label{fig:theory1}
\end{figure}

 The 
 structure of Ca$_3$Ti$_2$O$_7$  consists of slabs that contain two layers of CaTiO$_3$ perovskite separated by CaO rocksalt layers.
At room temperature, Ca$_3$Ti$_2$O$_7$ crystallizes in the
orthorhombic polar space group $A2_1am$ \cite{Elcombe1991,Senn2015}.  This 
structure 
[Fig. \ref{fig:theory1}(a)] decomposes into three distinct 
distortions: an out-of-phase octahedral tilt  ($a^-a^-c^0$ in Glazer notation \cite{Glazer1972}) that
transforms like the $X_3^-$ irreducible representation  of the high-symmetry prototype structure 
$I4/mmm$, an in-phase octahedral rotation 
($a^0a^0c^+$ in Glazer notation) that transforms
as $X_2^+$, and a polar distortion that transforms like
$\Gamma_5^-$. Each distortion is represented by a structural order parameter with amplitude $Q$ and phase $\phi$ \cite{Nowadnick2016}. 
For the chosen setting of the orthorhombic axes relative to the tetragonal axes 
the twin  domains 
are labelled by different settings of the space group symbol: $A2_1am$ (twin A) and $Bb2_1m$ (twin B). 
Figure  \ref{fig:theory1}(b) shows their structure. 
In twin A, the $X_3^-$ tilt axis (and  polarization direction) lie along [-110], whereas  they lie along [110] in twin B. The $X_2^+$ rotation axis lies along [001] in both twins, but the adjacent perovskite slabs have different rotation senses: in twin A the rotations in the two slabs are in-phase (red arrows) whereas in twin B, they are out of phase. 
The boundaries connect domains with different spontaneous strain states, so mechanical compatibility conditions determine their orientation \cite{Tagantsev2010}. 
The wall symmetry is obtained by combining the symmetry elements of the two domains and those transforming one domain into the other. To go from twin A to B, all three structural order parameters  rotate (change phase $\phi$) by 90$^\circ$ [Fig. \ref{fig:theory1}(b, c)]. 
%
Thus, the $X_3^-$ tilt axis rotates by 90$^{\circ}$, whereas the sense of the $X_2^+$ rotation reverses in every other perovskite slab. 
%
%
At the center of the domain wall, the local structure  is $C2mm$ [Fig. \ref{fig:theory1}(b)], where the $X_3^-$ tilt axes in slabs 1 and 2 are perpendicular to each other (tilts about [100] and [010] axes in slabs 1 and 2, respectively). The  $X_2^+$ rotation is unchanged in slab 1 and  zero in slab 2. Thus, the amplitude of the $X_3^-$ order parameter remains relatively constant across the domain wall, whereas the $X_2^+$  order parameter amplitude is suppressed in the middle of the path [Fig. \ref{fig:theory1}(d, e)].
Whether local $C2mm$ structure is realized at the center  depends on wall width. If it is atomically thin, the structural change from one domain to the other will be abrupt, but if (as observed here) the walls are wide, a macroscopic region with $C2mm$ symmetry may be realized.

\subsection{Locating domain walls in different fields of view}

\begin{figure}[b!]
    \includegraphics[width=4.5in]{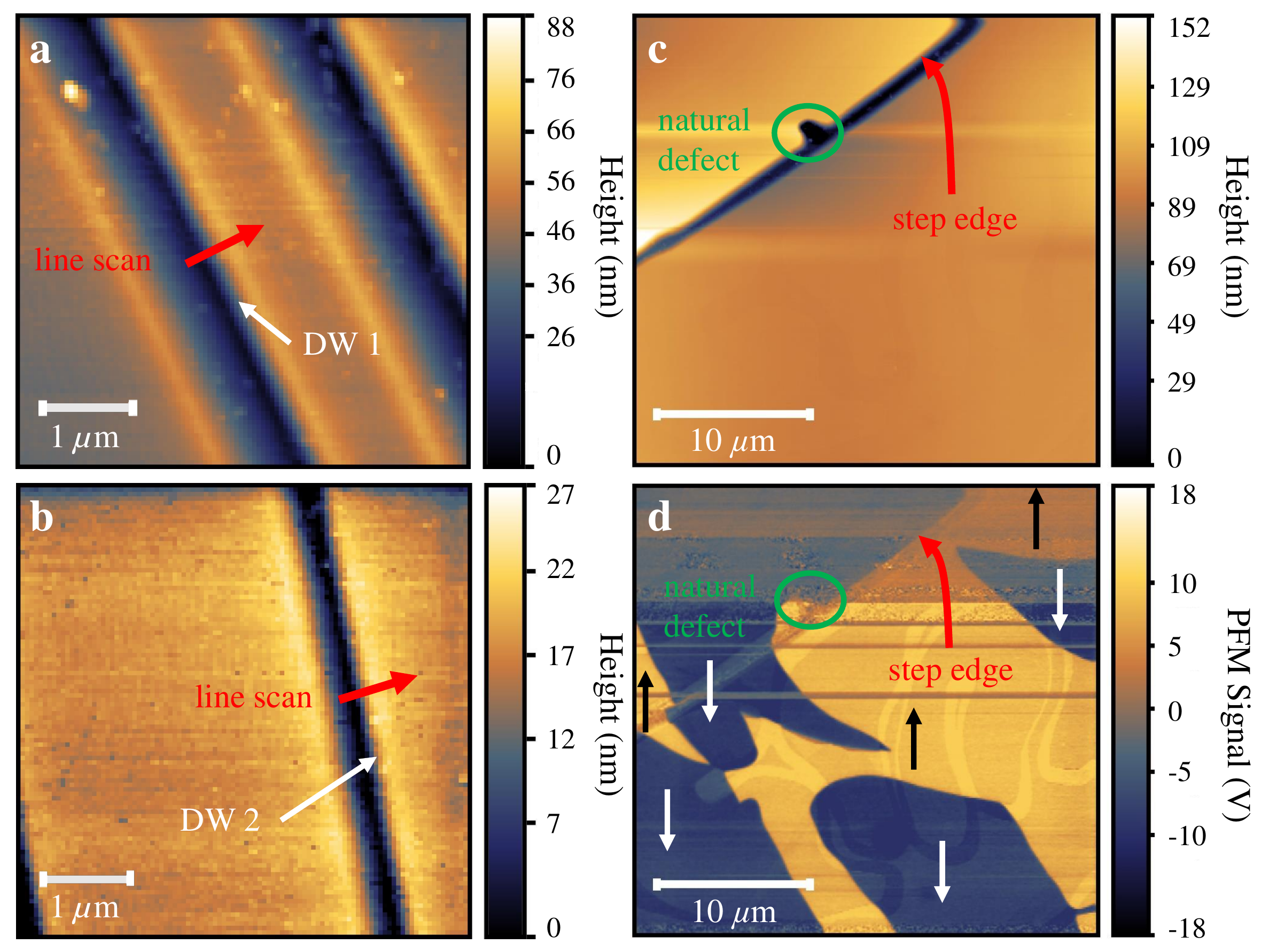}
    \caption{{{\bf Combining microscopy techniques to locate domain walls.}
    (a, b) Atomic force microscopy (AFM) images of the crystal surfaces showing the two ferroelastic domain walls  of interest (at the edges of the dark blue stripes). These ferroelastic walls separate domains of different spontaneous strain and are also 90$^{\circ}$ ferroelectric walls. DW 1 and DW 2 refer to domain walls 1 and 2. Red arrows indicate direction and path of the line scans. The nano-spectroscopic line scans are taken perpendicular to the wall, and the contact 
angle from one domain to another is 90$^{\circ}$. (c) AFM topography of a smooth area near an identified surface defect (indicated by a green circle) and step edge of approximately 100 nm height (indicated with a red arrow) compared with (d) the piezoresponse force microscopy (PFM) image of the same area revealing the placement and orientation of the 180$^{\circ}$ ferroelectric domains, indicated by yellow(+) or blue(-) regions with black or white arrows to indicate the polarization direction). All of these structures are present at room temperature.}}
    \label{Scanning}
    \end{figure}

Figure \ref{Scanning} summarizes the character of the different ferroelastic and ferroelectric domain walls in Ca$_3$Ti$_2$O$_7$ at room temperature. 
The ferroelastic (90$^{\circ}$ ferroelectric)  walls are apparent under cross-polarized light and readily identified by a color change  as linear twin boundaries in an optical microscope as illustrated in Supplementary Figure 1.
%
They appear as faint parallel lines in higher magnification images. 
These  features are present in a field of view where we have both atomic force microscopy (AFM) topography and near-field infrared spectroscopy.  Nano-spectroscopic line scans (where a spectrum is acquired at each pixel) are set up accordingly to cross ferroelastic walls of interest. The line scans discussed here are indicated with arrows in Fig.  \ref{Scanning}(a, b).  
%
%
We also identified candidate 180$^{\circ}$ ferroelectric walls for analysis and near-field line scans using a combination of AFM, piezoforce microscopy, and a careful examination of the ridges and topography of the crystal surface as illustrated in Fig. \ref{Scanning}(c, d).  These walls are much more challenging to locate because they meander between twin boundaries. A detailed explanation of this process and a summary of our spectroscopic findings are  available  in the Supplemental Information. 

\subsection{Near-field imaging of ferroelastic domain wall phonons}

\begin{figure}[tbh]
    \includegraphics[width=1\textwidth]{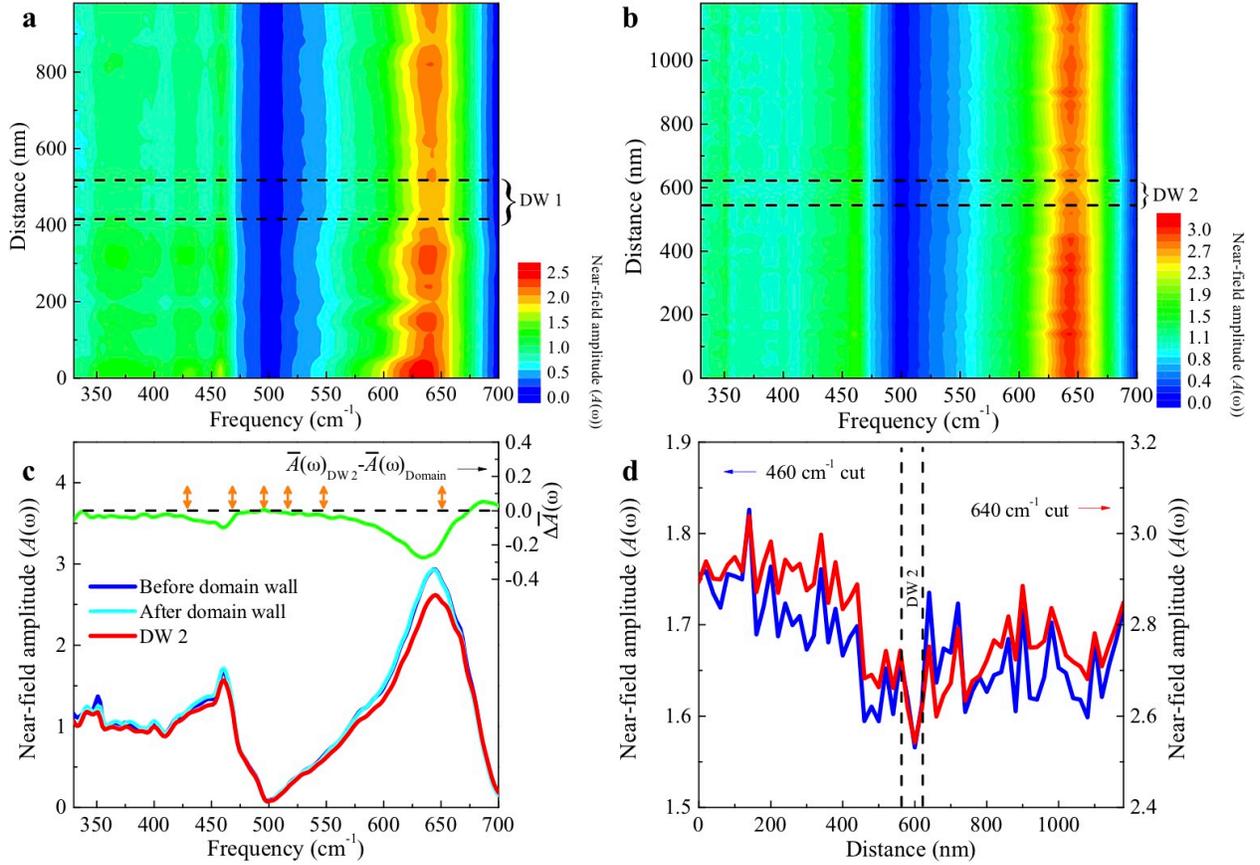}
    \caption {{\bf Near-field infrared spectroscopy of Ca$_3$Ti$_2$O$_7$.} (a, b) Contour plots of the near-field amplitude normalized to a gold reference across two different domain walls as  indicated in Fig. \ref{Scanning}(a, b). We label these walls as DW 1 and DW 2. The step size is 20 nm, and the tip resolution is 20 $\times$ 20 nm$^2$. The black dashed lines indicate the  domain wall locations, and the brackets  denote  effective wall widths. (c) Fixed distance cuts of the contour spectra in panel (b) show $A$(${\omega}$) at the ferroelastic domain wall compared with two different point scans away from the wall.  Average changes in the near-field amplitude, calculated as ${\Delta}{\bar{A}}$(${\omega}$) = ${\bar{A}}$(${\omega}$)$_{\textrm{DW 2}}$ - ${\bar{A}}$(${\omega}$)$_{\textrm{Domain}}$, reveal the difference and, at the same time, reduce the noise. Here,  the bars denote an average response. This analysis demonstrates that wall phonons have reduced amplitude and a slight blue shift. The 6 $A_1$ modes within the experimental energy window are indicated by double-sided orange arrows. (d) Fixed frequency cuts of the contour data in panel (b) showing how intensity at 460 and 640 cm$^{-1}$ varies across  DW 2. According to the calculation of ${\Delta}{\bar{A}}$(${\omega}$) in panel (c), these frequencies are most sensitive to the presence of the domain wall. They are also very near the calculated $A_1$ symmetry vibrational modes (shown as orange arrows in panel (c)).}
    \label{Realspace1}
    \end{figure}

Figure \ref{Realspace1}(a, b) displays the near-field scattering amplitude $A$(${\omega}$), which encodes the sample dielectric response across two different ferroelastic domain walls in the hybrid improper ferroelectric Ca$_3$Ti$_2$O$_7$. Each  contour plot shows the line scan distance  (the exact position of which 
is shown in Fig. \ref{Scanning}(a, b)) vs. frequency, with the color scheme indicative of near-field amplitude. 
We selected these particular scans to illustrate typical wall variations. 
The 
features in $A$(${\omega}$) are in reasonable agreement with the far field spectra (Supplemental Information).

A group theoretical analysis shows 72 zone-center phonons that transform as
\begin{equation}
\Gamma = 19A_1 \oplus 17A_2\oplus 19 B_1 \oplus 17 B_2.
\end{equation} 
The $A_1$, $B_1$, and $B_2$ modes are infrared and Raman active, whereas the $A_2$ modes are Raman active only. 
By considering the atomic displacement patterns arising from the phonon eigenvectors, the broad structures with maxima at   
645 and 450 cm$^{-1}$ are assigned as  Ti-O stretching and  Ti-O-Ti bending modes, respectively. In order to associate these localized vibrations with the rotations and tilts that are  crucial for the trilinear coupling  and ferroelectric switching, we projected out the character of each calculated phonon. This process is described in detail below.  We find that the Ti-O stretching mode has a component that transforms as  $X_2^+$, whereas the Ti-O-Ti bending mode transforms primarily as $X_3^-$. Thus, the symmetry of long-range rotations and tilts are mapped onto the more localized vibrations that are available in our experimental energy window.

 Strikingly, the near-field infrared spectrum is sensitive to the ferroelastic (90$^{\circ}$ ferroelectric) domain walls.  
  Focusing first on Fig. \ref{Realspace1}(a), we see that this line scan crosses a  
  twin boundary.  
 Both spectral amplitude and lineshape are altered across the wall. 
  This is interesting and important because domain walls in perovskites are traditionally considered to be atomically sharp boundaries \cite{Cao1990}. 
Turning to the second set of data in Fig. \ref{Realspace1}(b), a near-field  scan over an independent ferroelastic domain wall again reveals a significant decrease in phonon amplitude and line width. There are also very slight frequency shifts (toward the blue) that are at the limit of our resolution.
We therefore see that while the height and width of the wall  vary 
somewhat, 
the general spectral characteristics are similar.

Point spectra taken from the contour data [Fig. \ref{Realspace1}(c)] unveil a more traditional spectral view of the ferroelastic wall which we can compare to that of the surrounding  domain. 
In order to highlight spectral changes between the wall and the surroundings, we calculated a difference spectrum as  ${\Delta}{\bar{A}}$(${\omega}$) = ${\bar{A}}$(${\omega}$)$_{\textrm{DW 2}}$ - ${\bar{A}}$(${\omega}$)$_\textrm{{Domain}}$. This quantity, which reveals average changes (indicated by overbars) in the near-field amplitude  is shown in the upper part of Fig. \ref{Realspace1}(c).  
%
We find  that the wall phonons have reduced amplitude and a slight blue shift. The frequency shift is at the limit of our sensitivity, which unfortunately precludes a more detailed analysis. In any case, the blue shift suggests that there is a spontaneous strain across the wall \cite{Tagantsev2010}.   The strain across the wall occurs because a ferroelastic wall connects domains with different spontaneous strain states; this strain together with the modulation of order parameter amplitudes determines the local wall structure. 
%
Another property of interest is conductivity. 
We immediately notice that the twin boundary is semiconducting rather than metallic because there are strong phonons with no hint of a Drude response. 
Therefore, the ferroelastic walls are not metallic in the conventional sense - although they may be slightly  more conducting
than their surroundings.  To verify this observation, we calculated the band gap for the  bulk $C2mm$ structure (the hypothetical structure realized at the midpoint of the domain wall). We find that the computed gap is 0.26 eV less than that of the 
$A2_1am$ structure - not even close to closing the 3.94 eV gap in this system \cite{Cherian2016}. 
This confirms that the ferroelastic walls remain insulating.

Constant frequency cuts in the range of the Ti--O stretching and Ti--O--Ti  bending modes  
uncover another surprising aspect of the walls [Fig. \ref{Realspace1}(d)]. We take these cuts of the contour data at 460 and 640 cm$^{-1}$, where 
the change in near field amplitude between the wall and the surrounding domain as quantified by ${\Delta}{\bar{A}}$(${\omega}$) in Fig. \ref{Realspace1}(c) is strongest. 
%
Rather than an atomically sharp boundary, this direct and microscopic probe of the ungerade modes suggests that there is significant width to the structural distortion. 
We find widths from 60 to 100 nm for the range of domain walls investigated.  
This is akin to the length scale of structural relaxations in strained epitaxial thin films. 
The four different types of ferroelastic domain walls in Ca$_3$Ti$_2$O$_7$ (head-to-head, head-to-tail, tail-to-head, and tail-to-tail) \cite{Oh2015} along with the prediction from Landau theory that charged walls are thicker than their neutral counterparts \cite{Yudin2015} provide a natural explanation for this variation.

\subsection{Relating near-field amplitude to the order parameters}

We now  consider how 
localized phonons map onto the underlying long-range structural order parameters in Fig. \ref{fig:theory1}(d, e).
%
We know that the 19 $A_1$ phonons maintain $A2_1am$ crystal symmetry across the wall, so we can loosely think of the $A_1$ phonons as excitations of the structural order parameters. 
Since distortions that transform like the $X_3^-$, $X_2^+$, $\Gamma_5^-$, and $\Gamma_1^+$ irreducible representations of $I4/mmm$  contribute to the $A2_1am$ structure,  each $A_1$ phonon can (in principle) excite a mixture of these four structural order parameters. 
We therefore performed a change of basis and projected the computed $A_1$ phonon eigenvectors ${\bf e}_{A_1}$ onto a basis of symmetry adapted modes ${\bf e}_{i\tau}$ that transform like the irreducible representation $\tau=\{\Gamma_1^+, \Gamma_5^-, X_3^-, X_2^+\}$ of the high-symmetry prototype structure $I4/mmm$. 
Figure ~\ref{Phonons}(a) displays the results of this projection. Details are in the Supplemental Information.

 Remarkably, certain phonons almost completely overlap with a single symmetry adapted mode--exciting only one structural order parameter. Whereas, others are a mixture. Focusing first on the low-frequency phonons with calculated frequencies 428 and 466 cm$^{-1}$, we find that they transform primarily as ${\bf e}_{X_3^-}$. These phonons are Ti-O-Ti bends although they involve different bond angles [Fig.~\ref{Phonons}(b, c)]. We next consider the phonons calculated to be at 652 and 546 cm$^{-1}$. They primarily overlap with ${\bf e}_{\Gamma_1^+}$ and ${\bf e}_{X_2^+}$. The atomic displacement patterns [Fig.~\ref{Phonons}(d, e)] reveal that the 652 cm$^{-1}$ phonon is a $c$-polarized Ti-O stretch, whereas the 546 cm$^{-1}$ mode is an $ab$-polarized Ti-O stretch. This motivates our assignment of the broad experimental structure centered at 645 cm$^{-1}$ as bond stretching. The width of the spectral peak suggests that the $ab$-plane and $c$-directed Ti-O stretching modes overlap.


  There are three distinct types of oxygen centers in the layered Ruddlesden-Popper structure.  These include the equatorial oxygens O$_{\textrm{eq}}$, as well as two different types of apical oxygens: those that border the rocksalt layer O$_{\textrm{RS}}$, and those that lie in the middle of the perovskite slab, O$_P$. We find that the 466 cm$^{-1}$ phonon is primarily a Ti-O$_P$-Ti  bend, whereas the 428 cm$^{-1}$ mode involves both Ti-O$_{\textrm{RS}}$-Ti and Ti-O$_{\textrm{eq}}$-Ti bending motion.
From this analysis, we conclude that the broad experimental peak centered near 645 cm$^{-1}$  contains excitations that transform like $X_2^+$  and   $\Gamma_1^+$, whereas the wide experimental feature centered near 450 cm$^{-1}$ in the near field spectrum 
transforms primarily as 
the $X_3^-$ 
irreducible representation. 

\subsection{Order parameter trends vs. near-field response of ferroelastic domain walls}

In order to test the correspondence between these symmetry objects, we plotted the near-field amplitude of the 460 and 640 cm$^{-1}$ phonons as a function of distance across the ferroelastic domain wall and overlaid predictions for how the order 
parameters change across the wall. 
Importantly, these two frequencies are most sensitive to the presence of the domain wall, and they are very near the calculated $A_1$ mode positions [Fig. \ref{Realspace1}(c)]. 
%
%
The agreement, while not perfect, has several striking aspects. 
Our model predicts that the $X_3^-$ amplitude 
is relatively constant across the wall whereas the
amplitude of the $X_2^+$ rotation changes significantly and is suppressed in the center.
%
Domain wall 1 (DW 1) exhibits reasonable overall agreement with these predictions [Fig. \ref{Comparison}(a, b)]. 
The  460 cm$^{-1}$ feature remains relatively constant across the wall, 
although the anticipated minimum in the fixed frequency scan at 640 cm$^{-1}$  is not well pronounced. 
%
%
%
Domain wall 2 (DW 2) is different [Fig. \ref{Comparison}(c, d)], illustrating what we have found to be typical variations. 
The fixed frequency near-field scan at 460 cm$^{-1}$ is relatively flat  across the  the wall - in agreement with the predictions of the  $X_3^-$ order parameter. 
At the same time, 
 the   640 cm$^{-1}$ feature is  suppressed at the center of the wall, as anticipated.
%
%
%
%
Overall, both $X_3^-$ and $X_2^+$ track the behavior of the walls fairly well albeit with some deviation. 
Therefore,  we can loosely but not completely think of the long-range rotations and tilts as mapping onto the more localized vibrations that are available in our experimental energy window. 
%
%
Mixing effects [Fig. \ref{Phonons}(a)] and signal-to-noise issues  are the primary reasons that the agreement is not better.

This order parameter framework provides appealing insight into the spatial extent of the ferroelastic domain walls in Ca$_3$Ti$_2$O$_7$. 
Twin boundaries are traditionally considered to be quite narrow \cite{Cao1990,Salje2012}, although recent work suggests they may be wider than previously supposed \cite{Schiaffino2017}. 
In systems with octahedral rotations, the width and energy of a ferroelastic wall depends on the orientation of the octahedral rotation axes with respect to the domain wall plane \cite{Xue2014,Cao1990}. 
 In particular, symmetry constraints that require the octahedral rotation amplitude to go to zero at the center of the wall can increase the wall's width. 
%
%
%
In Ca$_3$Ti$_2$O$_7$, as the $X_2^+$ order parameter rotates by 90$^\circ$ across the ferroelastic wall, symmetry dictates that the $a^0$$a^0$$c^+$ rotation turns off in every other perovskite slab [Fig. \ref{fig:theory1}(c)]. 
%
%
The spatial extent of the ferroelastic walls - ranging from from 60 to 100 nm in our measurements - may therefore originate from frustration of the $X_2^+$ rotation. 
In addition to the octahedral rotation order parameters discussed here, the spontaneous strain also changes across the ferroelastic wall.  This means that the elastic properties also play a role in determining the  domain wall width and energy \cite{Tagantsev2010,Cao1990_2}.  Based on experimentally reported room temperature lattice parameters \cite{Senn2015}, Ca$_3$Ti$_2$O$_7$ has a small orthorhombic distortion with spontaneous strain $\eta=(b-a)/(b+a) = 8\times10^{-4}$. 
The   
 elastic energy per area of a ferroelastic wall can be estimated as $\eta^2 C d$, where $C$ is the elastic stiffness coefficient and $d$ is the domain wall width. Taking $C$ from DFT calculations \cite{Lu2016}  and estimated wall widths from our experiments, we find the elastic energy per area to be roughly 10 mJ/m$^2$. This relatively small elastic energy contribution may make wide walls a favorable situation. 

\section{Discussion}

In this work, we combine synchrotron-based infrared nano-spectroscopy and theoretical modeling to unlock a nanoscale view of ferroelastic domain walls in hybrid improper ferroelectric Ca$_3$Ti$_2$O$_7$. While domain walls have long been known to play a key role in determining the functionality of ferroic materials, exploration of the atomic- and nano-scale structure and properties of these walls is in its early stages, enabled by new high-resolution imaging techniques and complementary theoretical methods. In particular, we elucidate how the phonon response evolves across a 90$^{\circ}$ ferroelastic wall and relate it to the underlying structural changes that occur within the wall. We find that these twin boundaries  have a surprisingly large spatial extent, 
suggesting that ferroelastic walls are not always narrow objects.

\section{Methods}

\paragraph{{\bf Crystal growth and scanning techniques to locate domain walls:}} 
High quality single crystals of Ca$_3$Ti$_2$O$_7$ were synthesized
by  optical floating zone techniques as described previously \cite{Oh2015}.   Surface
topography was scanned using  atomic force microscopy, and
ferroelastic domain walls were identified using a combination of
visual inspection of orthorhombic twin boundaries and
cross-polarized light. We also carried out piezoforce microscopy to
identify ferroelectric domains and walls. Identification of common step edges and 
defects allow these images to be overlaid - even though they are not in the same field of view. In this manner, we are able to navigate around the surface and examine the lattice dynamics across different ferroelastic and ferroelectric domain walls.

\paragraph{{\bf Synchrotron-based near field infrared spectroscopy:}} 
Near-field infrared
spectroscopy was performed using the setups at beamlines 5.4 and 2.4 at the
Advanced Light Source, Lawrence Berkeley National Laboratory 
\cite{Bechtel2014}.  The apparatus is configured as an asymmetric Michelson interferometer, in which one arm consists of an AFM (Bruker Innova or Neaspec neaSNOM) and the other arm is a moving mirror.  Synchrotron infrared light is focused and scattered off an AFM tip in close proxmity to the sample surface.  The scattered light is combined with the reference beam from the moving mirror on a silicon or KRS-5  beamsplitter and detected by a liquid helium cooled Ge:Cu detector. Scanning the mirror of the reference arm creates an interferogram, which is Fourier-transformed to obtain 
both amplitude and phase information, which is related to the real and imaginary parts of the optical dielectric function of the material. The incident light is $p$-polarized with respect to the sample such that the polarization is parallel to the tip axis.  This configuration enables the most efficient coupling to the antenna modes of the tip, but there is also a component of polarization in the plane of the sample due to the focusing angle of the off-axis parabolic mirror.  However, the strong enhancement of the metal tip localizes and enhances the optical field linearly polarized parallel to the tip-axis, such that the technique is most sensitive to phonon and vibrational modes perpendicular to the sample surface \cite{Muller2016}. To discriminate the near-field signal from the far-field background signal, the AFM is operated in non-contact (tapping) mode at a typical frequency of 250-300 kHz, and the corresponding detector signal is demodulated at the twice the tapping frequency. 
Our measurements were performed with typical free oscillation tip amplitudes in the 80-100 nm range with a setpoint between 70-78\% of the free-tapping amplitude, such that the engaged tapping amplitude values were in the 55-75 nm range.   Routine approach curves with free oscillation tip amplitudes in the 80-100 nm range taken on gold indicate a near-field tip enhancement within $<$30 nm of the surface with little to no second harmonic signal when the sample is withdrawn further.  
Different setpoints within the 70-78\% range have minimal to no effects on the sample amplitude or shape. The setpoint is typically chosen to have the least damping (i.e. closer to 78\%) while still maintaining good AFM feedback and reproducibility between the forward and backward traces.  This approach minimizes tip wear as well .

Our near-field infrared measurements were carried out at room temperature
over a frequency range between  330 and 800 cm$^{-1}$. The near field spectra are corrected for the limited transmissivity of the entire setup by normalizing the power spectrum of the sample to that of a gold reference 
mirror. Both second and third harmonic signals were analyzed. The spatial resolution of this technique is limited by the AFM tip radius, which is typically $<$25 nm for the tips used (Nanosensor PtSi-NCH). Our line scans employed a step size of 20 nm. Multiple line scans across the same domain wall are very similar in character with variances on the order of the noise level. Scans across different domain walls reveal some variations which is why we show the results for two different walls in this work.

\paragraph{{\bf Far field infrared spectroscopy:}} 
We also
carried out traditional, far-field spectroscopy using a suite of
Fourier transform and grating spectrometers covering the spectral range from  15 to 65,000 cm$^{-1}$ for comparison with the near-field infrared response. Data was collected in reflectance mode in the $ab$-plane and along the $c$-direction. 
A summary of mode positions and assignments is available in Supplementary Information.

\paragraph{{\bf Symmetry analysis and lattice dynamics calculations:}}
Density functional theory calculations were performed using
projector augmented wave (PAW)
pseudopotentials \cite{Kresse1999} and the PBEsol
functional, as implemented in
VASP \cite{Kresse1993}.  We used the (Ca\_sv, Ti\_sv, O) VASP
pseudopotentials with electronic configurations 3$s^2$3$p^6$4$s^2$,
3$s^2$3$p^6$4$s^2$3$d^2$, and 2$s^2$2$p^4$, respectively.
Calculations were performed in a 48 atom supercell commensurate with
both orthorhombic twins of Ca$_3$Ti$_2$O$_7$ with a $6\times 6\times
2$ Monkhorst-Pack $k$-point mesh and a  600 eV plane wave cutoff.
All structural relaxations were performed with a force convergence
tolerance of 2 meV/$\AA$. The theoretical curves across domain walls were obtained by considering a path through the bulk energy surface connecting the two domains and calculating a sequence of bulk structures along that path. 
This was done using nudged elastic band calculations, allowing the ions to relax at each image along the path -  between the $A2_1am$ domain to the midpoint structure $C2mm$  and into the $Bb2_1m$ domain.
Note that we are considering a sequence of bulk structures along the path through the bulk energy landscape that connects the two domains; we are not taking into account order parameter gradients and strains, which are known to be important at the domain wall. While this level of theory does not permit quantitative predictions of, e.g., domain wall widths, we find that it places a simple focus on the evolution of the structural order parameters across the domain wall. Phonon frequencies and eigenvectors were calculated using density functional perturbation theory \cite{Baroni2001}. We performed group theoretic analysis
with the aid of ISOTROPY \cite{Stokes} and the Bilbao Crystallographic Server \cite{Aroyo2011, Aroyo2006, Aroyo2006_2, Kroumova2003} and visualized crystal
structures using VESTA \cite{Momma2008}. 
%

\section{Acknowledgments}

Research at the University of Tennessee is supported by the U.S.
Department of Energy, Office of Basic Energy Sciences, Materials
Science Division under award DE-FG02-01ER45885.
Portions of the measurements utilized beamlines 5.4 and 2.4 at the Advanced Light Source, which is a DOE Office of Science User Facility operated under contract no. DE-AC02-05CH11231, including the remote user program from NSLS-II under contract DE-SC0012704. Work at Cornell is supported by the Army Research Office under grant W911NF-10-1-0345. 
Work at Rutgers is funded by the Gordon and Betty Moore Foundation’s EPiQS Initiative through Grant GBMF4413 to the Rutgers Center for Emergent Materials. EAN acknowledges additional support from the New Jersey Institute of Technology. M. B. R. and O. K. achknowledge support from the U. S. Department of Energy, Office of Basic Sciences, Division of Material Sciences and Engineering, under award no. DE-SC0008807. O. K. acknowledges support from the ALS Postdoctoral Fellowship program. Work at Postech is supported by the National Research Foundation of Korea (NRF) funded by the Ministry of Science and ICT (No. 2016K1A4A4A01922028).

\section{Competing Interests}
The authors declare no competing interests as defined by Nature Research. 

\section{Author Contributions}
This project was conceived by JLM and KAS.  Crystals were 
 grown by BG, CJW, and SWC, and piezeoforce microscopy was carried out by SJL and SWC. Spectroscopic measurements were performed
by KAS, SF, SNN, OK, GLC, MCM, HAB, and JLM. Data manipulation and visualization assistance was provided by JKK. First-principles calculations and predictions based on a symmetry analysis of the order parameters were carried out by EAN and CJF.  All authors discussed the data. The manuscript
was written by KAS, EAN, HAB, and JLM, and all authors commented on the document.

\section{Data Availability}
Relevant data is available upon request from the corresponding
authors, Jan Musfeldt (email: musfeldt@utk.edu) and Elizabeth Nowadnick (email: enowadnick@ucmerced.edu)

\section{References}

\bibliographystyle{apsrev}

\begin{figure}[tbh]
    \includegraphics[width=6.0in]{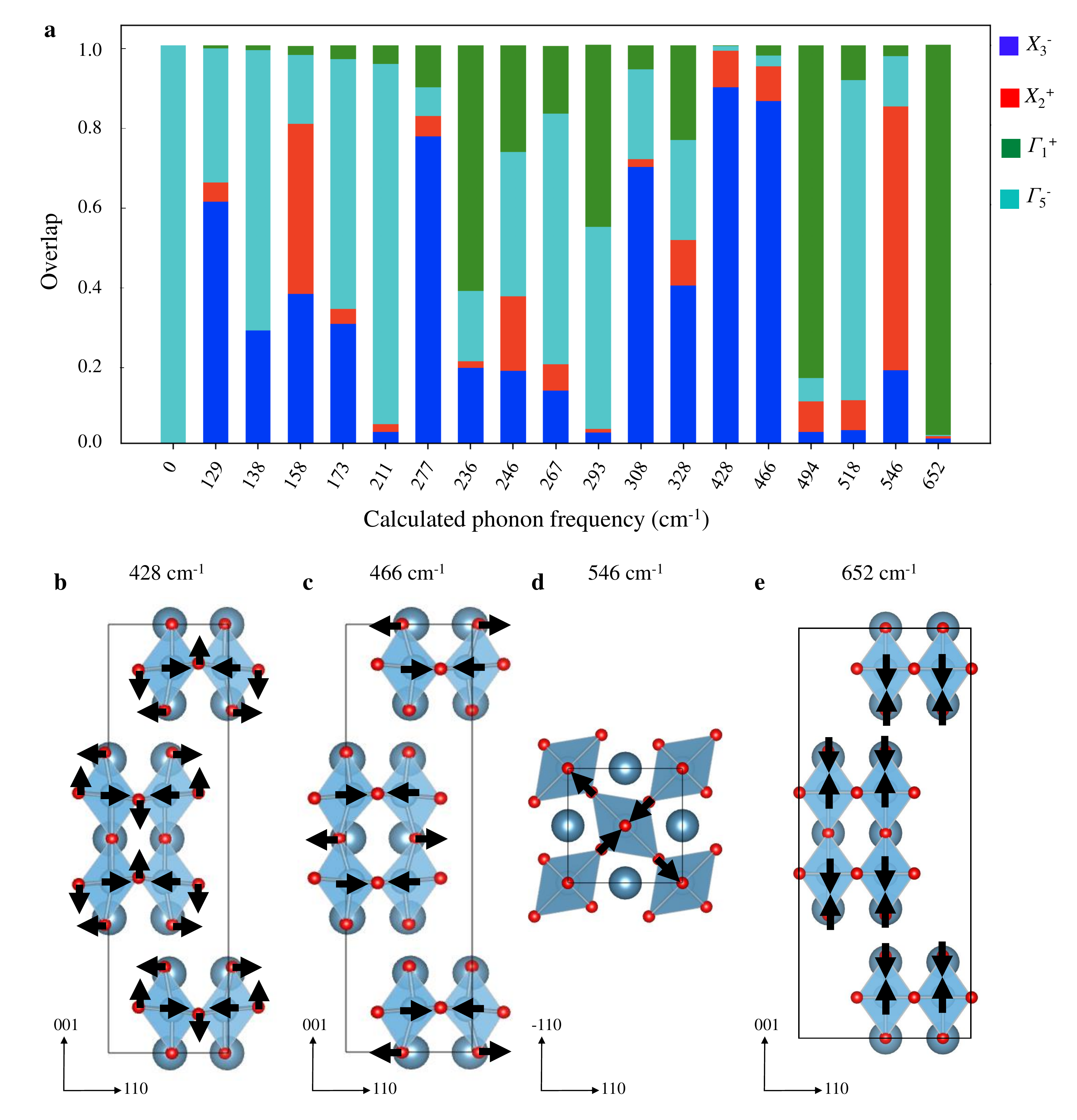}
    \caption{{{\bf Phonon overlaps and calculated displacement patterns.} (a) The 19 $A_1$ symmetry phonon eigenvectors projected onto the $I4/mmm$ symmetry adapted modes. The labels are the calculated phonon frequencies.  (b-e) Atomic displacement patterns of four selected phonons from (a). The black arrows display the largest atomic motions within each displacement pattern. The 428 cm$^{-1}$ and 466 cm$^{-1}$ phonons largely consist of Ti-O-Ti bond angle bends. The 546 cm$^{-1}$ and 652 cm$^{-1}$ phonons are bond-stretching modes in the $ab$ plane and along the $c$-axis, respectively.}}
    \label{Phonons}
    \end{figure}

\begin{figure}[tbh]
    \includegraphics[width=6.5in]{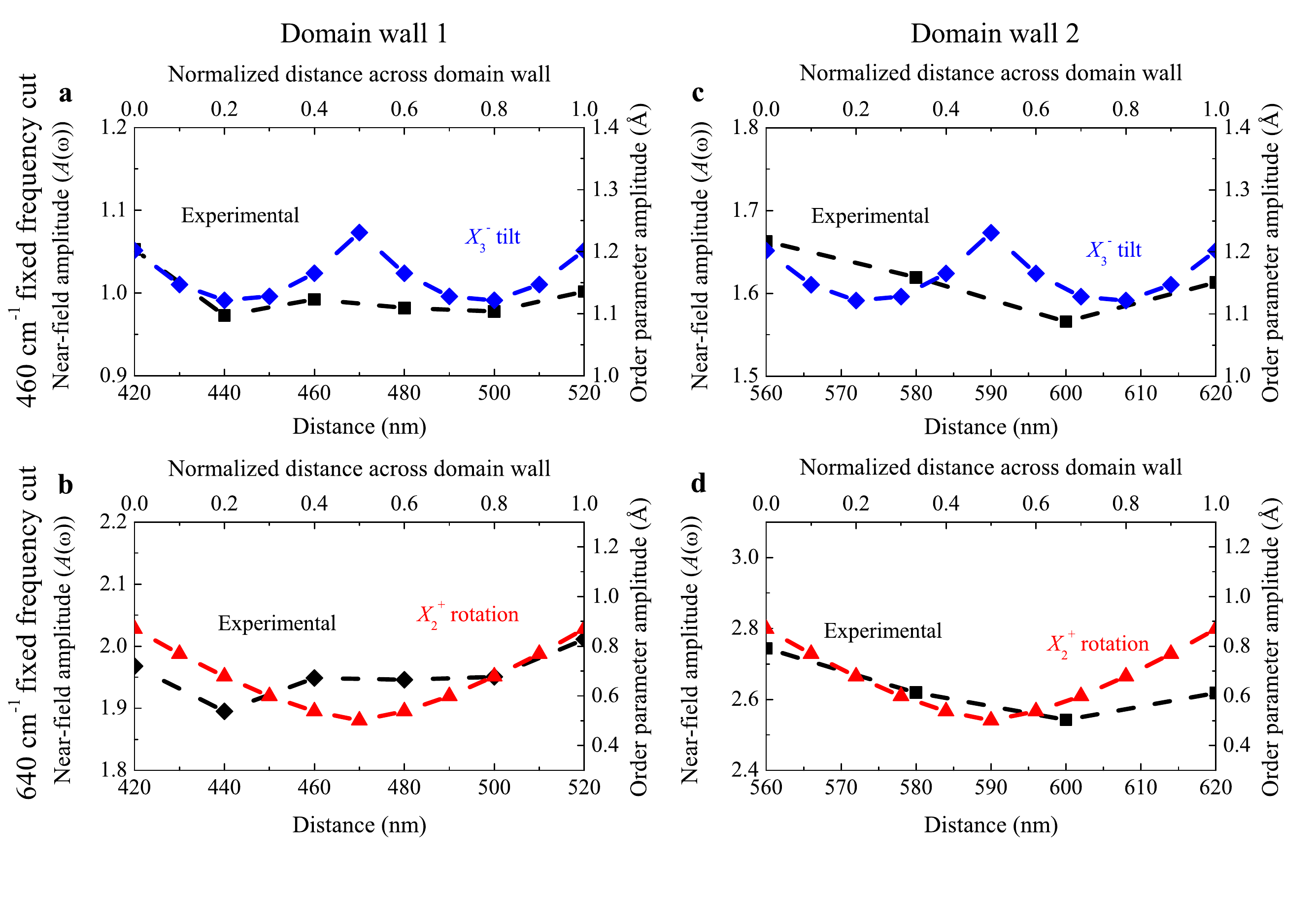}
    \caption{{\bf Order parameter trends across the structural domain wall.} (a-d) Comparison of predicted  $X_3^-$ (tilt, blue diamonds) and $X_2^+$ (rotation, red triangles) order parameters with the measured near-field amplitude at 460 and 640 cm$^{-1}$ across the two walls of interest from Fig. 3(a, b). Both the normalized distance across each wall (treated here as a scalable parameter) as well as the actual line scan positions (determined from the near-field response) are shown. The microscopic distortions across the wall relax more slowly 
    than topography would suggest. 
   %
  %
   }
    \label{Comparison}
    \end{figure}

\end{document}